\documentclass[twoside]{ilcws08}
\usepackage[latin1]{inputenc}
\usepackage{graphicx}
\usepackage{epsfig}
\usepackage{color}
\usepackage{wrapfig}
\usepackage{amssymb,amsmath,array}

\begin{document}
\title{Resolution Studies on Silicon Strip Sensors \\ with Fine Pitch } 
\author{S. Haensel$^1$, T. Bergauer$^1$, Z. Dolezal$^2$, M. Dragicevic$^1$, Z. Drasal$^2$, \\ M. Friedl$^1$, J. Hrubec$^1$, C. Irmler$^1$, W. Kiesenhofer$^1$, M. Krammer$^1$, P. Kvasnicka$^2$
\vspace{.3cm}\\
1 - Institute of High Energy Physics of the Austrian Academy of Sciences (HEPHY) \\
Nikolsdorfergasse 18, 1050 Vienna - Austria
\vspace{.1cm}\\
2 - Faculty of Mathematics and Physics, Charles University \\
V Holesovickach 2, 180 00 Prague - Czech Republic\\
}

\maketitle

\begin{abstract}
In June 2008 single-sided silicon strip sensors with 50~$\mu$m readout pitch were tested in a highly energetic pion beam at the SPS at CERN. The purpose of the test was to evaluate characteristic detector properties by varying the strip width and the number of intermediate strips. The experimental setup and first results for the spatial resolution are discussed.
\end{abstract}

\section{Introduction} 

\begin{wrapfigure}[22]{r}{0.38\columnwidth}
\vspace{-35pt}
\centerline{
\includegraphics[width=0.37\columnwidth]{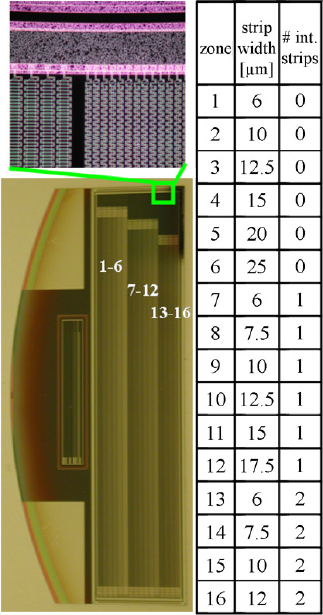}}
\vspace{-10pt}
\caption{Sensor layout}
\label{Fig:sen}
\end{wrapfigure}

The goal of this study is to evaluate the ideal strip geometry of silicon strip sensors with 50~$\mu$m readout pitch in terms of spatial resolution. This readout pitch was chosen because it is the smallest pitch feasible for large scale module production needed for future silicon trackers. Therefore, a single-sided, multi geometry silicon strip sensor was designed. The varying parameters, strip width and number of intermediate strips, influence the strip capacitance and the charge sharing ability of the strips, both basic properties that contribute to the spatial resolution of the sensor. With these sensors, an experiment was performed at CERNs SPS accelerator using 120~GeV pions.

\section {Sensor layout}

Eight multi-geometry silicon strip sensors, manufactured by Hamamatsu Photonics, Japan, were used for this study. The sensors are single sided with a thickness of 320~$\mu$m and an active area of 15$\times$64~mm$^{2}$. A bulk resistivity of 6.7~k$\Omega$cm ensures full sensor depletion at a voltage below 100~V. The readout sensor area is divided into 16 zones with various strip geometries, separated by a gap of one missing strip (Fig.~\ref{Fig:sen}). Each zone consists of 16 AC-coupled readout strips of 50~$\mu$m pitch, which are biased using integrated 20~M$\Omega$ poly-silicon resistors. These resistors (small insert in Fig.~\ref{Fig:sen}) have different lengths, dependent on the number of intermediate strips, which can easily be recognised by the length of the aluminium strips in the sensor picture. The strips in the different zones have varying strip widths and a different number of intermediate strips (0, 1 or 2).

\section{Sensor tests}
 
\begin{wrapfigure}[12]{r}{0.43\columnwidth}
\vspace{-32pt}
\centerline{
\includegraphics[width=0.45\columnwidth]{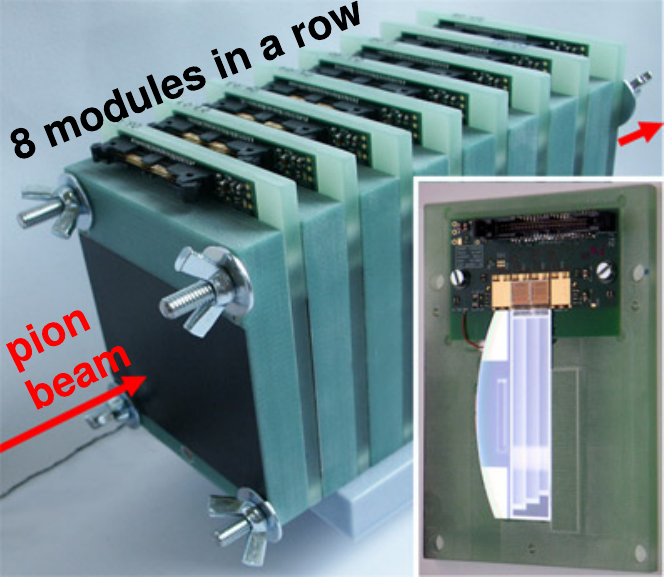}}
\vspace{-10pt}
\caption{Stack of modules and module layout}
\label{Fig:mod}
\end{wrapfigure}

Prior to module assembly all sensors were electrically tested. CV measurements revealed full depletion voltages at an average of 60~V. IV-curves measured up to 800~V showed stable behaviour far beyond the depletion voltage. The interstrip capacitance rises linearly with increasing strip width, because the distance between the strips decreases, with a varying offset for the different number of intermediate strips.

\section{Detector modules}

For the test beam a stack of eight identical detector modules, with a distance of 2~cm between the silicon planes, was used (Fig.~\ref{Fig:mod}). Each detector module is build of an Isoval\textregistered11 support frame which holds the sensor and the front end hybrid with two APV25 readout chips connected to the 256 sensor channels. A stack of eight modules was used to increase statistics and enable autonomous tracking, analising one sensor while using the remaining seven as ``telescope".

\section{Data acquisition}

The readout system, a prototype for the upgrade of the Belle Silicon Vertex Detector, was also developed in Vienna. The APV25 readout chips, connected to the sensor strips, deliver the data via an electrical cable to the Repeater Boards (REBO). The data gets digitalized and processed on two 9U VME Boards. To synchronise the telescope DAQ and our readout system, a Trigger Logic Unit (TLU) was used to centrally distribute event number and the trigger signal from scintillators in front of and behind the sensors. A PC running a $\rm LabWindows^{TM}$/CVI application stored the data and enabled online monitoring. Most of the data were taken in raw mode with a trigger rate of approximately 50~Hz during the beam spills.

\section{Test beam}

\begin{wrapfigure}[9]{r}{0.45\columnwidth}
\vspace{-40pt}
\centerline{
\includegraphics[width=0.44\columnwidth]{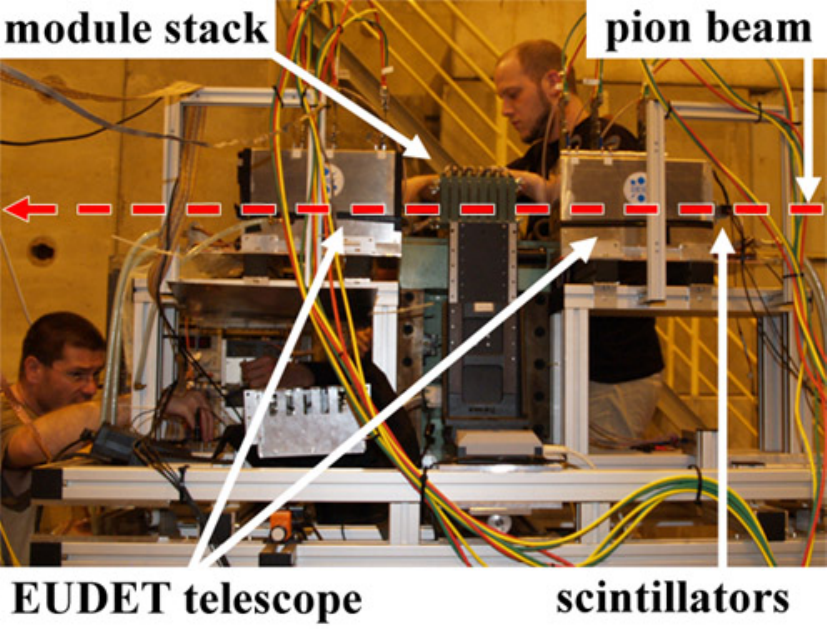}}
\vspace{-10pt}
\caption{Test beam setup}
\label{Fig:setup}
\end{wrapfigure}

The test beam took place at the SPS-H6B area at CERN. Our module stack was mounted on a xyz-rotation-stage between the two times three CMOS pixel sensors of the EUDET telescope\cite{EUDET-telescope} (Fig.~\ref{Fig:setup}). The SPS provided 120~GeV Pions on a beam spot of 4$\times$2~cm$^2$ during spills of 5~s with intervals of 20-40~s. The beam intensity was reduced to a minimum by narrowing collimators in the beam line in order to minimize multiplicity in the telescope.

\section{Measurements}

Three different sets of runs were performed:

1) Three runs were taken to determine the spatial resolution of the different strip geometries. Since the telescope's pixel sensors (7~mm) are more narrow than our sensors (13~mm), the resolution run had to be performed three times with shifted module positions (100k events each). A malfunction of the xyz-rotation-stage lead to a 0.3° rotation of our module stack w.r.t. the beam axis. 

2) A high voltage scan was performed to examine the dependence of the detector signal on the sensor operation voltage. 13 runs with bias voltages between 10~V and 200~V were performed (10k events each).

3) During an angle scan we took data for six different angles between the beam line and the sensor planes (90° to 30° - 10k events each).

\section{Preliminary results}

\begin{wrapfigure}[17]{r}{0.48\columnwidth}
\vspace{-45pt}
\centerline{
\includegraphics[width=0.47\columnwidth]{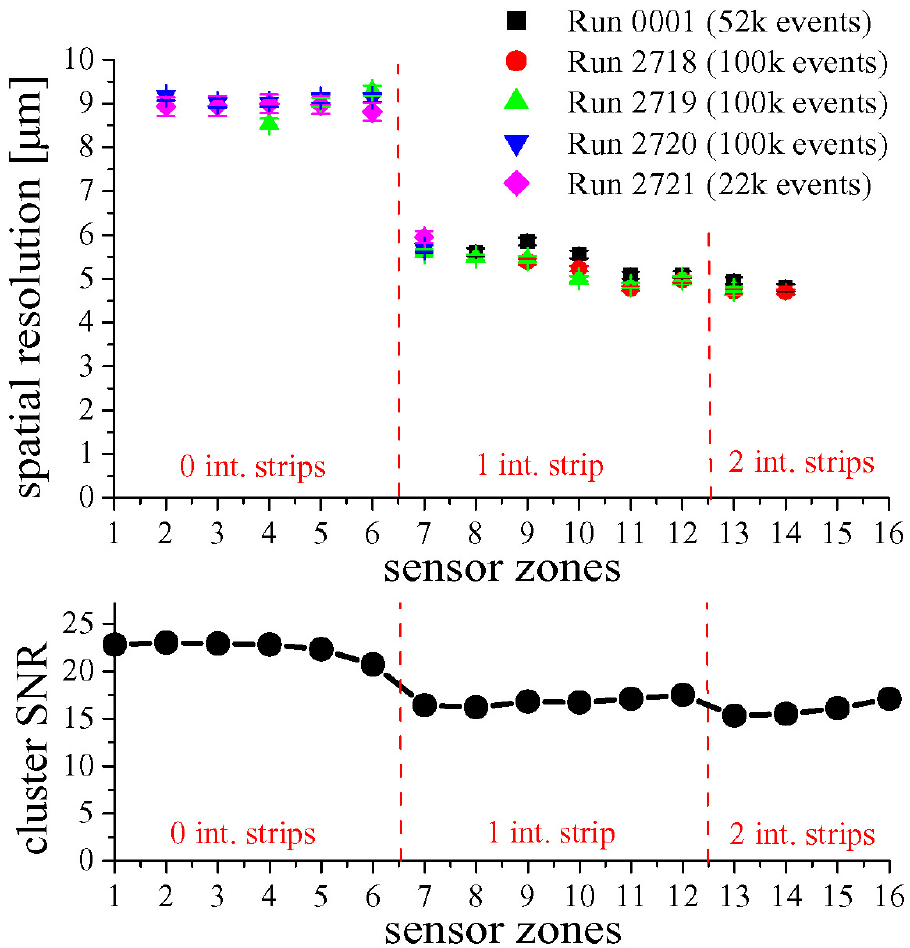}}
\vspace{-10pt}
\caption{top: preliminary spatial resolutions; bottom: cluster SNR}
\label{Fig:resolutions}
\end{wrapfigure}

These preliminary results only show the data from resolution runs, excluding the telescope data. The spatial resolution is calculated by iteratively estimating the residuals of a single sensor plane in the centre of the module stack, using the other seven strip sensors as reference ``telescope". Only tracks with hits in all eight sensors are considered. Since the module stack was rotated by 0.3° during these runs, too little statistics of the edge zones 1, 15 and 16 remain (excluded in Fig.~\ref{Fig:resolutions}). The spatial resolution of the zones with zero intermediate strips is around 9~$\mu$m, relatively independent of the strip width. Zones with one intermediate strip have a spatial resolution of 5-6~$\mu$m, an improvement of about 40~\%. For zones with two intermediate strips this preliminary analysis shows a resolution below 5~$\mu$m. For comparison, the digital resolution for a strip pitch of 50~$\mu$m is about 14~$\mu$m ($\sigma_x(digital)=(readout~pitch)/ \sqrt{12}$). The cluster signal to noise ratio:
$SNR=(most~propable~signal) / [(average~strip~noise)\times \sqrt{(average~cluster~width)}]$
is almost constant for zones with the same number of intermediate strips, only slightly rising with the strip width (Fig.~\ref{Fig:resolutions} bottom). The strip noise is almost constant for all zones with an average of 770 electrons. The mean cluster width slightly increases with the number of intermediate strips, but is always around 2. Comparing the average SNR of zones with zero (22.44) and one (16.79) intermediate strips shows a decrease of about 25~\%. More detailed analysis are ongoing.

%



\begin{footnotesize}

\end{footnotesize}


\end{document}